\pdfoutput=1

\documentclass[11pt]{article}

\usepackage{EACL2023}

\usepackage{times}
\usepackage{latexsym}

\usepackage[T1]{fontenc}

\usepackage[utf8]{inputenc}

\usepackage{microtype}

\usepackage{inconsolata}

\usepackage{graphicx}
\usepackage{lipsum}

\newcommand{\Kaucus}{\textbf{Kaucus}}

\title{\texttt{\textcolor{darkblue}{KAUCUS}}: Knowledge Augmented User Simulators for Training Language Model Assistants}

\author{Kaustubh D. Dhole\\
Department of Computer Science \\
  Emory University\\
  Atlanta, USA \\
  \texttt{\textcolor{darkblue}{kdhole@emory.edu}} \\
  }

\begin{document}
\maketitle
\begin{abstract}
An effective multi-turn instruction-following assistant can be developed by creating a simulator that can generate useful interaction data. Apart from relying on its intrinsic weights, an ideal user simulator should also be able to bootstrap external knowledge rapidly in its raw form to simulate the multifarious diversity of text available over the internet. Previous user simulators generally lacked diversity, were mostly closed domain, and necessitated rigid schema making them inefficient to rapidly scale to incorporate external knowledge. In this regard, we introduce~\Kaucus{}, a~\textbf{K}nowledge-\textbf{Au}gmented~\textbf{U}ser~\textbf{S}imulator framework, to outline a process of creating diverse user simulators, that can seamlessly exploit external knowledge as well as benefit downstream assistant model training. Through two GPT-J based simulators viz., a~\textbf{Retrieval Augmented Simulator} and a~\textbf{Summary Controlled Simulator} we generate diverse simulator-assistant interactions. Through reward and preference model-based evaluations, we find that these interactions serve as useful training data and create more helpful downstream assistants. We also find that incorporating knowledge through retrieval augmentation or summary control helps create better assistants. 
\end{abstract}

\begin{figure*}
\centering
\fbox{
  \includegraphics[width=\linewidth]{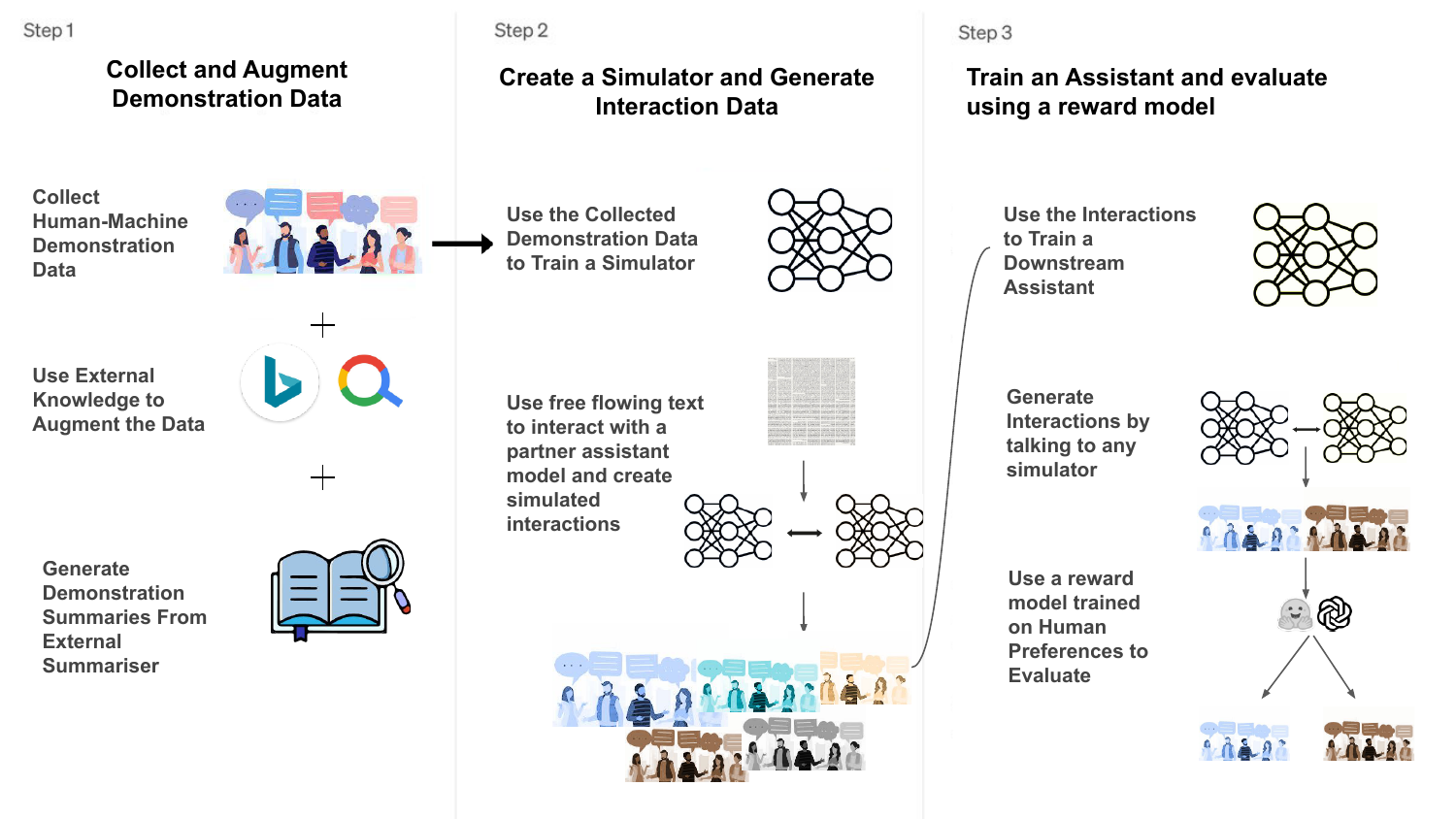}}
  \caption{The complete three step framework of~\Kaucus{} -- creating, utilizing and evaluating a user simulator.}
  \label{fig:teaser}
\end{figure*}

\section{Introduction}

Significant advancements in Large Language Models (LLMs) have made them exceptionally adept in conversational applications like virtual assistants~\cite{touvron2023llama,fitzgerald2022alexa,openai2023gpt4, team2023gemini}. This proficiency is largely attributed to the notably parallelizable transformer architecture~\cite{vaswani2017attention} enabling these models to utilize extensive pre-training datasets effectively~\cite{2019t5,together2023redpajama}.
To create effective assistants, LLMs are then further enhanced by learning from human interactions including popular paradigms such as RLHF~\cite{bohm-etal-2019-better,ziegler2019finetuning,NEURIPS2022_b1efde53}. Such conversational human alignment of assistants requires large amounts of interactive dialog data, both for training as well as testing. 

However, interactive data collection is a manual and slow process, particularly (a) for covering a wide range of user behaviors as well as (b) for diverse adversarial and behavior testing.

These challenges can be mitigated by simulating user behaviors by automating the generation of interactive data, reducing both time and cost, while maintaining control over the interactions. Simulated interactions can be executed at a much faster pace than manual collection efforts, limited only by the speed of inference. 

Yet, current user simulators lack diversity, are mostly closed domain, and require rigid schema for control or conversation grounding. The necessity of intermediate schema in the form of a knowledge base~\cite{kim2023soda} or handcrafted rules (like user persona or specific behaviors) while being excellent drivers to ground conversations, make it hard to develop scalable simulators -- that can utilize natural text freely available on the internet and rapidly create corresponding assistant models. A simulator should be able to exploit external knowledge rapidly and also be controllable without a rigid schema. We argue that such a knowledge simulator can be helpful in two ways -- It can seamlessly convert free-form text to useful training data without user intervention as well as provide a natural control to direct simulators for specific behaviors~\cite{mille2021automatic,cheng-etal-2023-compost}. 

Hence, in this work, we propose~\Kaucus, a~\textbf{K}nowledge~\textbf{Au}gmented~\textbf{S}imulator Framework\footnote{pronounced like Caucus derived from Algonquian cau'-cau'-as'u meaning ‘adviser'}. Through this framework, we demonstrate the usage of external sources of knowledge -- viz. Retrieval Augmentation and Summary Control -- for creating user simulators that can incorporate free-flowing text and result in better assistant training. 

The paper is organized as follows: In Section~\ref{related_work}, we first discuss existing work related to user simulators. In Section~\ref{general_framework}, we define simulators and introduce~\Kaucus, through two knowledge simulators. We further describe the efficacy of each through training and evaluating downstream assistant models. Our retrieval augmented simulator,~\textbf{SRAG} shows how retrieving relevant passages with a simple BM25 retriever can be used to improve intrinsic metrics as well as provide useful training data to train helpful assistants. We also introduce the summary-controlled setting,~\textbf{SCTRL} to build scalable simulators to exploit freely available text and further measure their performance with and without retrieval.

\section{Related Work}\label{related_work}
User simulators have been studied in various settings.~\citet{aher2023using} create four simulators that elicit behavior to judge an assistant’s fairness, rationality, grammaticality, and general knowledge, and then measure them qualitatively. Their simulators are models with different prompt templates. Training multi-agent interactions has been a popular choice in reinforcement learning.~\citet{horton2023large, argyle2023out} create simulations for economic purposes by endowing GPT3~\cite{brown2020language} with demographic characteristics and then get responses in various scenarios that match what is seen empirically.~\citet{irving2018ai} in AI safety has proposed using self-play and self-debate to train AI agents to pursue human goals and preferences. Two tasks in the collaborative benchmark, BIG-Bench~\cite{srivastava2023beyond} evaluate the model’s ability for self-evaluation by simulating specific human professions. They make the models to act as lawyers, tutors\footnote{\href{https://github.com/google/BIG-bench/tree/main/bigbench/benchmark_tasks/self_evaluation_tutoring}{BIG-Bench Self Evaluation Tutoring}}, judges\footnote{\href{https://github.com/google/BIG-bench/tree/main/bigbench/benchmark_tasks/self_evaluation_courtroom}{BIG-Bench Self Evaluation Courtroom}}, students, etc. and then have separate model instances to evaluate the conversation. Each of the roles is invoked by user-specific prompts like ``You are a lawyer'' and a subsequent model-based evaluation is performed by prompting to seek numerical ratings.

~\citet{kreyssig-etal-2018-neural}'s Neural User Simulations involve training encoder-decoder RNNs on dialogues between real users and a spoken dialogue system (SDS) in a restaurant domain and then using the trained simulator to train the policy of a reinforcement learning based SDS. They further use~\citet{schatzmann2005quantitative}'s cross-model evaluation to compare user simulators by training different policies with each simulator and testing it with other simulators.~\citet{gur-etal-2018-dialsql} encode dialog history and a goal to generate user responses for task-oriented dialog.~\citet{ds1,ds2} prompt LLMs with task-oriented dialog data, such as goals, and perform intrinsic evaluation over the generated data to show the effectiveness of their approaches.~\citet{kim2023soda} generate conversations grounded on common sense by prompting InstructGPT with knowledge base triples. Their human evaluations show that oftentimes humans prefer model outputs against their human-written counterparts.~\citet{liu2023one} leverage multiple user simulators to train task-oriented dialog systems.~\citet{faltings-etal-2023-interactive} utilize user simulators that offer edits to guide the model towards achieving a specified target text training them using Imitation Learning.

Other studies augment simulators with emotions~\cite{emotionsim} and trusting behaviours~\cite{kraus2023development}. For instance,~\citet{emotionsim} simulate user emotions alongside user behavior based on the user goal, the dialogue history, and persona.~\citet{giabbanelli2023gpt} utilize GPT-based models for scientific simulations while~\citet{schaefer2023large} explore LLMs to simulate biological systems.

With the popularity of large language models deployed in closed-source settings, bootstrapping training data from them has become useful.~\citet{alpaca} create downstream assistant models by training LLama-7B and 13B models~\cite{touvron2023llama} on 52K single-turn instruction following demonstrations generated through self-instruct~\cite{Self-Instruct} from~\texttt{text-davinci-003}~\cite{brown2020language}.~\citet{ChatAlpaca} create a dialog corpus by extending the same to the multi-turn setting.~\citet{dai2022dialoginpainting} show improved conversation retrieval by proposing a mechanism to convert Wikipedia passages to dialog.

On the other hand, retrieval augmentation has been the focus of many recent efforts~\cite{schick2023toolformer, sirenssurvey,wang-etal-2023-self-knowledge,li2022survey} as it offers advantages such as up-to-date information access beyond an LLM's training dataset, incorporation of proprietary or domain-specific data at runtime, and enhanced factuality in outputs compared to standard LLMs. Studies have been performed by training RAG systems end-to-end~\cite{Guu2020REALMRL, lewis2020retrieval} as well as using retrieval in context for various tasks~\cite{incontextralm,jiang-etal-2023-active,gao2023rarr,genqrensemble}.

\section{The~\Kaucus{} Framework}\label{general_framework}
In this section, we introduce~\Kaucus{}, a 3-stage framework, and outline the process of creating knowledge-augmented simulators as shown in Figure~\ref{fig:teaser}. Our approach involves the following steps:
\subsection{Data Collection and Augmentation} We start by gathering interaction data -- essentially conversations between a user and a base assistant LLM, which will be later augmented to enrich the training process. For instance, the base LLM could take the form of closed-source instruct models such as OpenAI’s GPT-4, Claude, or BingChat which are widely used for work.  

\subsection{Training a Language Model (LM) as a Simulator} The next step involves training a Language Model (LM) to act as a simulator. This LM can then serve as a conversation generator for data augmentation~\cite{dhole2023nl} or be integrated into a pipeline that relies on conversation interactions, such as Reinforcement Learning from Human Feedback (RLHF)~\cite{ziegler2019finetuning, ouyang2022training}. Our work focuses on the former. 
\subsection{Leveraging the User Simulator} Once the user simulator is trained, there are several methods to utilize for improving an assistant Language Model (LM). Our work resorts to data augmentation, which will be the focus of our second set of experiments. It involves using the user simulator to generate additional training data to enhance the assistant LM's performance.

\subsection{Evaluation}
To evaluate the effectiveness of the user simulator, we will employ both intrinsic and extrinsic metrics. Intrinsic metrics will be measured over the interactions with the simulator, assessing its performance in generating relevant and coherent responses. On the other hand, extrinsic metrics will be based on evaluating a downstream assistant model trained over these interactions, which will help us gauge the impact of the user simulator on overall assistant performance. We will describe the evaluation in detail in Section~\ref{evaluation}.

\begin{figure*}
\centering
  \includegraphics[width=\textwidth]{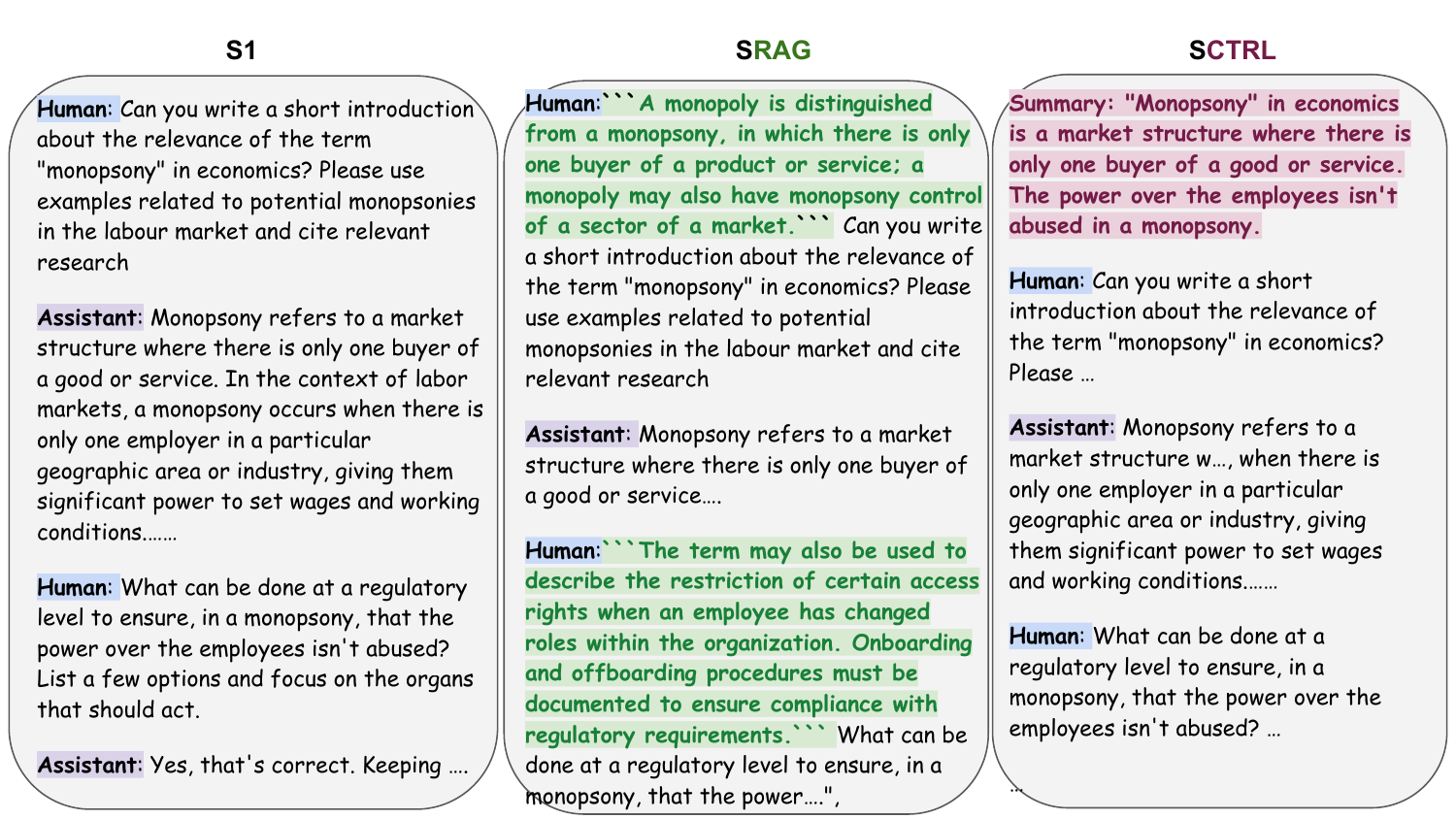}
  \caption{The format of the conversations used for training S1 (a vanilla simulator), SRAG (retrieved document shown in green), and SCTRL (summary shown in red).}
  \label{trainingdataformat}
\end{figure*}

\section{Methods and Experiments}\label{methods_and_experiments}
We now specifically describe the two types of knowledge-augmented simulators, viz. Utterance Grounded Simulators (S1 and SRAG) and Summary Controlled Simulators (SCTRL).

\subsection{Utterance Grounded Simulators}
Here we train simulators with human-machine demonstration data by feeding models the conversation history to create simulators that can be triggered from a starting utterance. We create two simulators -- S1 and SRAG by fine-tuning an unsupervised pre-trained GPJ-6B~\cite{gpt-j} model. We describe the training process for both below:

\subsubsection{S1} Simulator Trained on Anthropic and Open Assistant Conversations
\begin{itemize}
    \item \textbf{Training Data}: For training S1, we use demonstration data available through Open Assistant’s conversations~\cite{2023openassistant} and Anthropic’s helpful splits~\cite{bai2022training}. 
    \item \textbf{Format}: The Simulator’s training data consists of ~\texttt{(context, human-response)}
    pairs. For every ``Human'' utterance in all the conversations, we select all the previous utterances along with their speaker information and pass it as an input to the model. The input also consists of a ``Human:'' string at the end. The associated human response is passed as the output.
\end{itemize}

\subsubsection{SRAG} \textbf{Retrieval Augmented Simulator} Trained on Anthropic and Open Assistant Conversations with BM25 Retrieval on MSMarco

Simulators could benefit from the incorporation of external knowledge which can be helpful to steer the conversation, improve factuality and most importantly introduce variation. To test our hypothesis, we train the second simulator, SRAG by incorporating passages retrieved from an external retriever. 
\begin{itemize}
    \item \textbf{Training Data}: We augment the interactions used to train S1 with passage snippets from the MS-Marco dataset~\cite{nguyen2016ms, bajaj2016ms}, which is a large-scale dataset of 8.8M passages popularly used for information retrieval and reading comprehension. Having been generated from real users’ search queries, it provides a vast repository of documents collected on a plethora of topics over the web. 
    \item \textbf{Format}: We use ~\texttt{(context, human-response)} pairs in the style of S1 with human turns annotated with retrieved MSMarco passages. Using the human utterance as a query, we execute a BM25 retriever against an MSMarco Passage Index for every human turn. Each of the human utterances is then prepended with a retrieved passage as shown in Figure~\ref{trainingdataformat} in green. We use the MSMarco index provided by IRDatasets~\cite{irdatasets} and the BM25 implementation provided by PyTerrier~\cite{pyterrier}. 
\end{itemize}

\subsection{Summary Controlled Simulators}
The previous utterance-grounded setting relies on a conversational utterance at inference time to initiate the interaction. While it can be easy to obtain such conversational utterances using existing conversational datasets, they can quickly become scarce and out-of-date. It would be of interest to be able to scale over vast amounts of free text available over the web. However, most of the web data exists in a non-conversational format unsuitable for direct incorporation in the training process.

\textbf{SCTRL}: In that regard, we introduce the training of~\textbf{summary controlled simulators} that can utilize the conversational summary obtained from an external conversation summarizer during training. This can be potentially helpful in two ways -- It can provide a mechanism for the simulator to attempt to seamlessly convert ``free form text'' to ``interaction data'' while also coming up with the ``simulator trigger'' by itself reducing our reliance on conversational corpora. As compared to a fixed schema or a knowledge base, it can provide a natural control to guide simulators for specific behaviors via natural language texts which are generally available in plenty as compared to their conversational or interactive counterparts. 
\begin{itemize}
    \item \textbf{Training Data}: To create the training data, we append a conversational summary generated from an external conversational summarizer, at the beginning of the conversation. Our objective is to force the simulator to be able to learn the association between the initial non-conversational text and the subsequent conversation. We choose an existing ~\href{https://huggingface.co/knkarthick/MEETING-SUMMARY-BART-LARGE-XSUM-SAMSUM-DIALOGSUM-AMI}{BART Summariser}~\cite{wolf-etal-2020-transformers} fine-tuned on various dialog and non-dialog summarisation datasets like DialogSum, AMI and XSUM. 
    \item \textbf{Format}: We prepend the predicted summary at the start of the conversation as shown in Figure~\ref{trainingdataformat} in red.
\end{itemize}
We create the two summary-controlled counterparts of S1 and SRAG as SCTRL and SCTRL-RAG respectively.

\textbf{SCTRL-RAG} Summary Controlled Simulator Trained on Anthropic and Open Assistant Conversations with MSMarco BM25 Retrieval

We use a GPT-J-6B model RLHF fine-tuned on demonstration data as our base assistant model and our simulator. We use deepspeed~\cite{rasley2020deepspeed} to optimize training and train for 10 epochs on a learning rate of $10^{-6}$.

\section{Evaluation \& Results}\label{evaluation}
\subsection{Intrinsic Metrics}

We first seek to assess the ``diversity'' of the generated interactions. In assessing diversity, we utilize well-established reference-free lexical metrics viz. TTR, logTTR, RootTTR, HDD, and MTLD are based on type-token ratios and are quick to compute. The Measure of Textual Lexical Diversity (MTLD) is a prevalent and contemporary TTR metric that does not vary as a function of text length and explains textual information that similar lexical diversity approaches do not account for~\cite{mccarthy2010mtld}. It gauges the proportion of distinct word stems (types) to the overall word count (tokens). HDD is an alternative metric that captures additionally unique lexical information~\cite{mccarthy2010mtld}\footnote{Through a separate ancillary study, we also find that simulators trained on dialog data generate more diverse text as compared to pre-trained ones according to the above metrics.}.

We first generate 125 interactions by making each of the simulators interact with a fixed assistant model. The conversation is initialized with an existing Anthropic conversation in the case of S1 and SRAG and five more turns are generated (referred to as the augmented length). In SCTRL and SCTRL-RAG, 5 turns are generated from scratch from Anthropic's conversation summary. 
\begin{table*}[ht]
    \centering
    \resizebox{\textwidth}{!}{%
    \begin{tabular}{l|l|l|l}
        \textbf{Source} & \textbf{Type} & \textbf{Generated Interaction Data} & \textbf{Assistant} \\ \hline
        Human & -- & Assistant Trained With Anthropic\_8k & A0 \\ 
        S1 & Without Knowledge &Simulated Anthropic\_8k & A1 \\ 
        SRAG & With Retrieval Augmentation & Simulated Anthropic\_8k + MSMarco & A1-RAG \\ 
        S1-CTRL & With Summary Control & Simulated Anthropic\_8k*10 summaries & A1-CTRL \\
        S1-CTRL-RAG & Both & Simulated Anthropic\_8k*10 summaries + MSMarco & A1-CTRL-RAG \\ 
    \end{tabular}}
    \caption{The sources of various simulated data used in~\Kaucus{} to train the corresponding assistants}
    \label{table:assistant_data}
\end{table*}
We present the results in Table~\ref{table:lexical_diversity}. The metrics measure the lexical diversity only on the utterances generated via the simulator interaction (and not on the initial Anthropic conversation history that was fed to initiate the interaction). Across all metrics, incorporating a knowledge component, through retrieval augmentation (SRAG) or summary control (SCTRL) improves diversity. Incorporating both improves diversity across RootTTR and HDD metrics.

\begin{table}[!ht]
    \centering
    \resizebox{\columnwidth}{!}{%
    \begin{tabular}{l|ccccc}
        \textbf{Simulator} & \textbf{MTLD} & \textbf{Root TTR} & \textbf{LogTTR} & \textbf{HDD} \\ \hline
        \textbf{S1} & 23.177 & 2.918 & 0.818 & 0.04 \\ 
        \textbf{SRAG} & \textbf{24.632} & \textbf{3.223} & \textbf{0.82} & \textbf{0.134} \\ 
        \textbf{SCTRL} & \textbf{25.864} & \textbf{3.437} & \textbf{0.844} & \textbf{0.131} \\ 
        \textbf{SCTRL-RAG} & 22.761 & 2.976 & 0.766 & 0.278 \\ 
    \end{tabular}}
    \caption{Lexical diversity metrics on 125 conversations of each simulator. The top-2 highly diverse simulators are the knowledge-based ones - SRAG and SCTRL on all metrics.
}
\label{table:lexical_diversity}
\end{table}

\subsection{Extrinsic Metrics}
Although the aforementioned metrics can assist in evaluating and comparing various user simulators as potential data augmenters and generators, it is crucial to determine if they benefit subsequent assistant models. 
The RLHF paradigm, by training reward models, has demonstrated assistants that are more helpful, honest, and less harmful providing a promising direction for aligning with human preferences. In this regard, we resort to the family of reward and preference models to measure how well assistant models trained using data produced from various simulators perform.

\textbf{Training Downstream Assistant Models}: For each simulator trained (S1, SRAG,..), we create a subsequent assistant model (A1, ARAG, ...) and use reward modeling to measure the helpfulness of each of the assistant models. To create training data for each of the assistant models, we first simulate interactions between the corresponding simulator model along a separately held-out assistant model. 

For each utterance grounded simulator (S1 and SRAG), we use 8000 Anthropic conversations as triggers. Particularly, we utilize the complete Anthropic conversation as the starting history for both the simulator and the separately held-out assistant model and allow ten turns (5 pairs) of interactions to be generated. Using the simulator to generate longer contexts provides an opportunity to collect a larger number of~\texttt{(context, assistant-response)} pairs for training the downstream assistant model. 

For the retrieval augmented simulator, SRAG, it is necessary to retrieve passages relevant to the ongoing conversation. We hence use the previous assistant response as a query to our MSMarco Passage Index before generating the simulator turn. The top-ranked passage via BM25 is then placed at the end of the input to SRAG.

For generating interactions from SCTRL, we need free-flowing text as the initial trigger. We generate 8000 conversations from conversation summaries of the Anthropic dataset. We use additional 9*8K passages from MS-Marco as initial triggers to act as implicit summaries.

After generating the conversations, we convert them into~\texttt{(context, assistant-response)} pairs and use them as training data for predicting the assistant response given all the previous utterances. We call the subsequent assistant models A1, ARAG and ACTRL. The training details of each assistant model are described in Table~\ref{table:assistant_data}.

\textbf{Baseline}: We additionally train an assistant model, A0 using raw 8000 conversations from Anthropic to act as appropriate baseline.

\textbf{Test Set}: For evaluation, we utilize 200 utterances from the test set of Anthropic's dataset.

\textbf{FastChat Evaluation}: FastChat~\cite{zheng2023judging} is a platform for evaluating and serving LLMs. We resort to FastChat evaluation for prompting GPT-4~\cite{openai2023gpt4} for a comparative evaluation between two simulators. The process involves GPT-4 being input with two conversations, placed one after the other, along with an instruction to evaluate and generate a numerical score. We attribute a win, a loss, or a tie depending on whether the first (assistant model on the left in all the images) has a value greater, lesser, or equal to the second (one on the right). 

\textbf{SteamSHP Reward Model Evaluation}
SteamSHP-XL~\cite{pmlr-v162-ethayarajh22a} is a preference model fine-tuned on top of an instruction-tuned model FLAN-T5-XL~\cite{weifinetuned, longpre2023flan} to predict which response humans will find more helpful, given some context and two possible responses. On being prompted the same context, the reward model setting compares the probabilities assigned independently to each model response to infer the preference label.

\textbf{SteamSHP Preference Model Evaluation}~\cite{pmlr-v162-ethayarajh22a}
Preference modeling, like the FastChat Evaluation, compares two model responses through a single inference pass, which can be used to compute the probability of the first one being better than the second. 

To avoid any bias occurring through the order of two conversations, we also calculate the scores with the simulator order reversed in the prompt.

For each plot, the columns indicate the two assistant models being compared. The colors in blue for each row indicate when the evaluation system prefers the left-hand side model as compared to the right-hand side when compared against A0. 
\begin{figure}[ht]
\centering
\fbox{
  \includegraphics[width=\columnwidth]{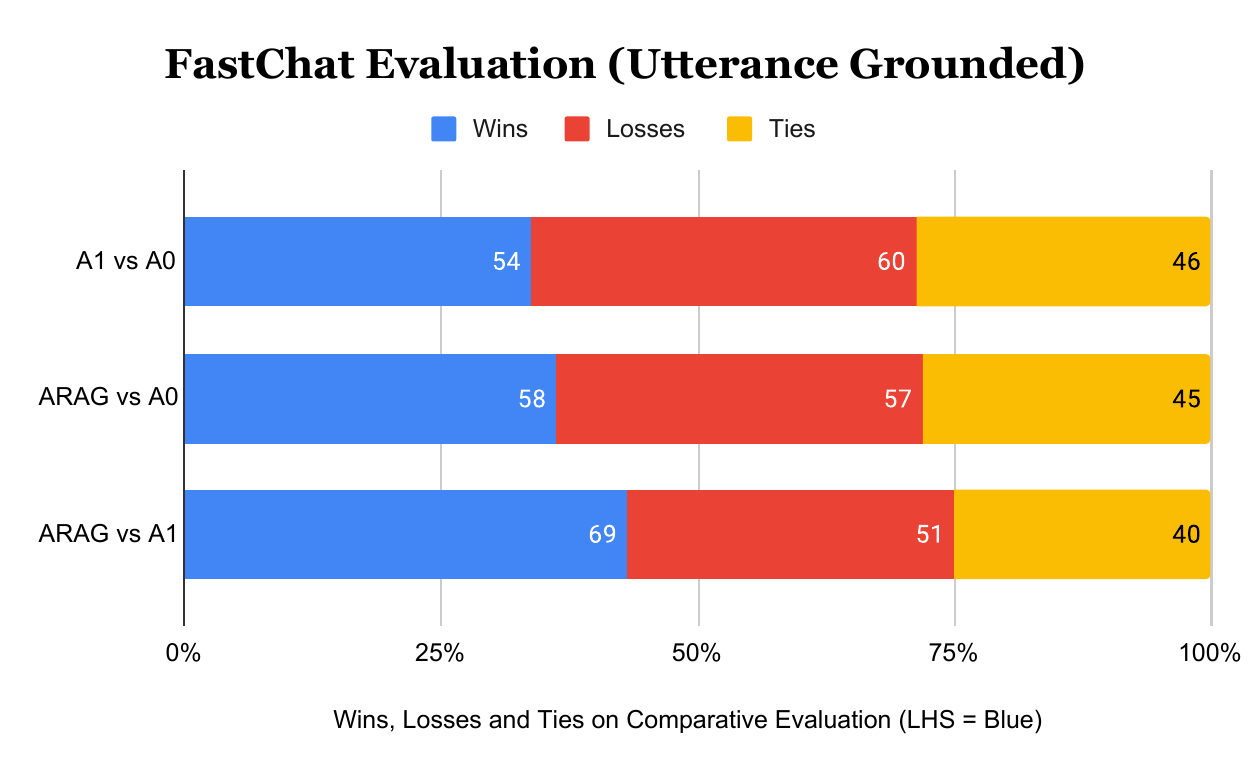}}
  \caption{FastChat Evaluation of Assistants created from Utterance Grounded Simulators (A1 and ARAG) against baseline assistant (A0)}
   \label{ug_fc}
\end{figure}
\begin{figure}
\centering
\fbox{
  \includegraphics[width=\columnwidth]{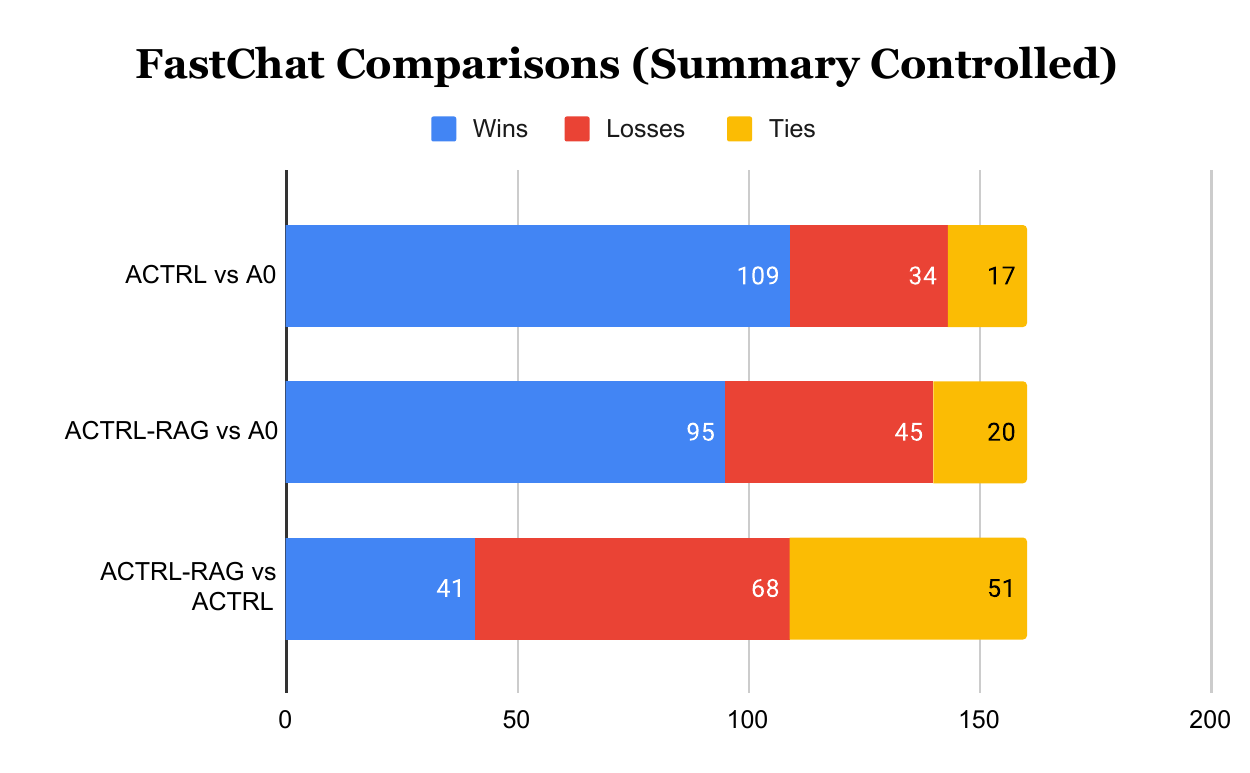}}
  \caption{FastChat Evaluation of Assistants created from Summary Controlled Simulators (-CTRL) against baseline assistant (A0)}
  \label{sg_fc}
\end{figure}
\begin{figure}[ht]
\centering
\fbox{
  \includegraphics[width=\columnwidth]{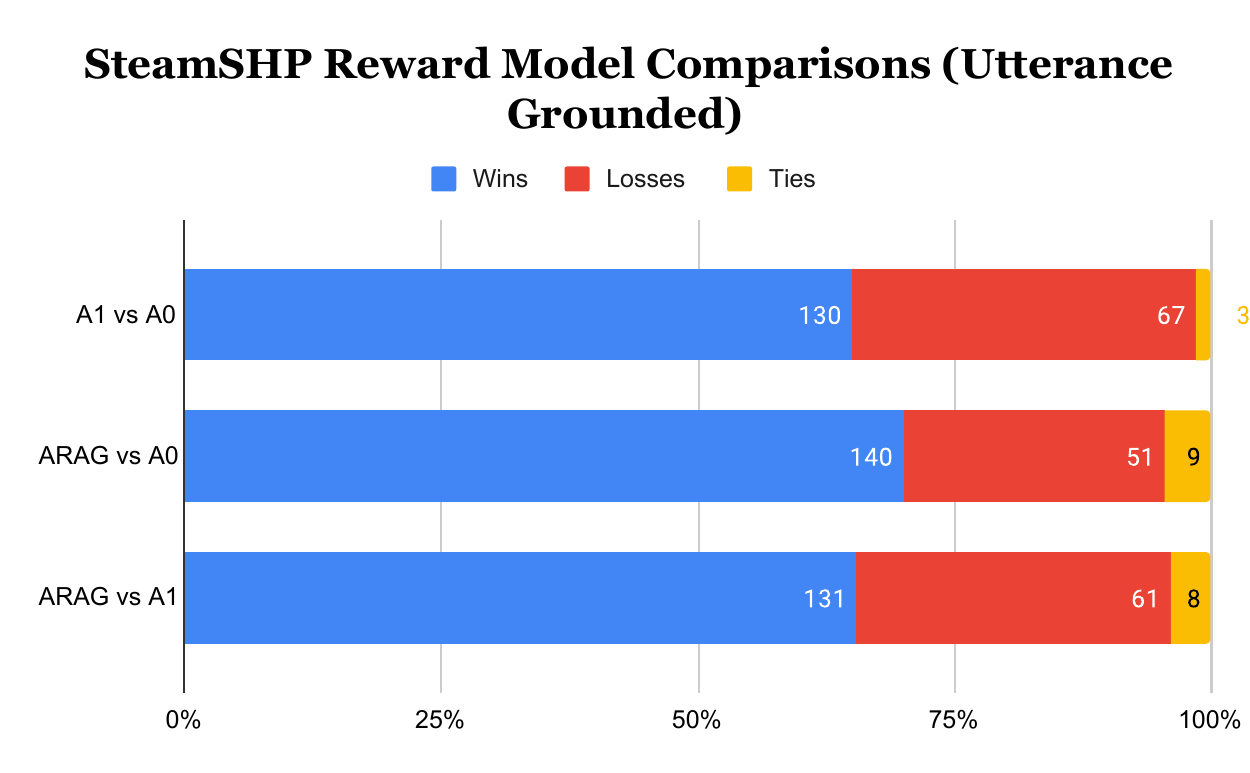}}
  \caption{SteamSHP reward model Evaluation of Assistants created from Utterance Grounded against baseline assistant (A0)}
  \label{ug_sr}
\end{figure}
\begin{figure}[ht]
\centering
\fbox{
  \includegraphics[width=\columnwidth]{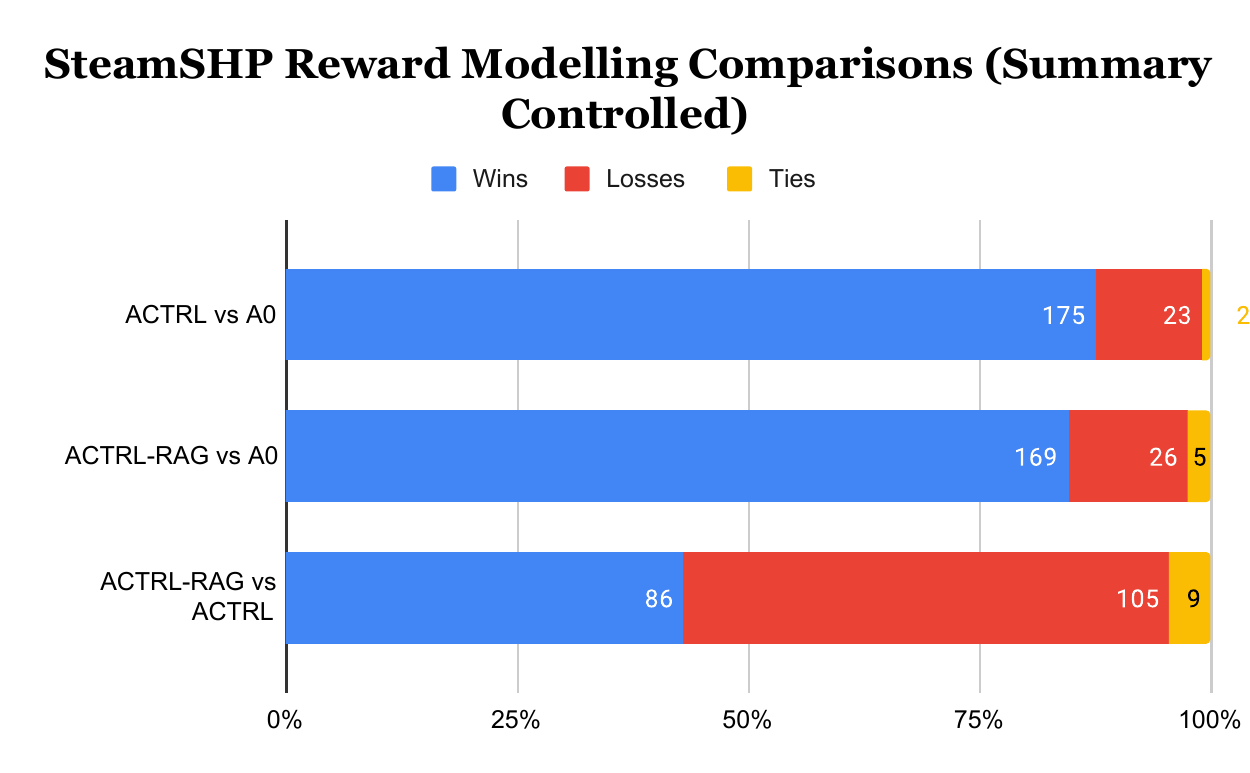}}
  \caption{SteamSHP reward model Evaluation of Assistants created from Summary Controlled Simulators (-CTRL) against baseline assistant (A0)}
  \label{sg_sr}
\end{figure}
\begin{figure}[ht]
\centering
\fbox{
  \includegraphics[width=\columnwidth]{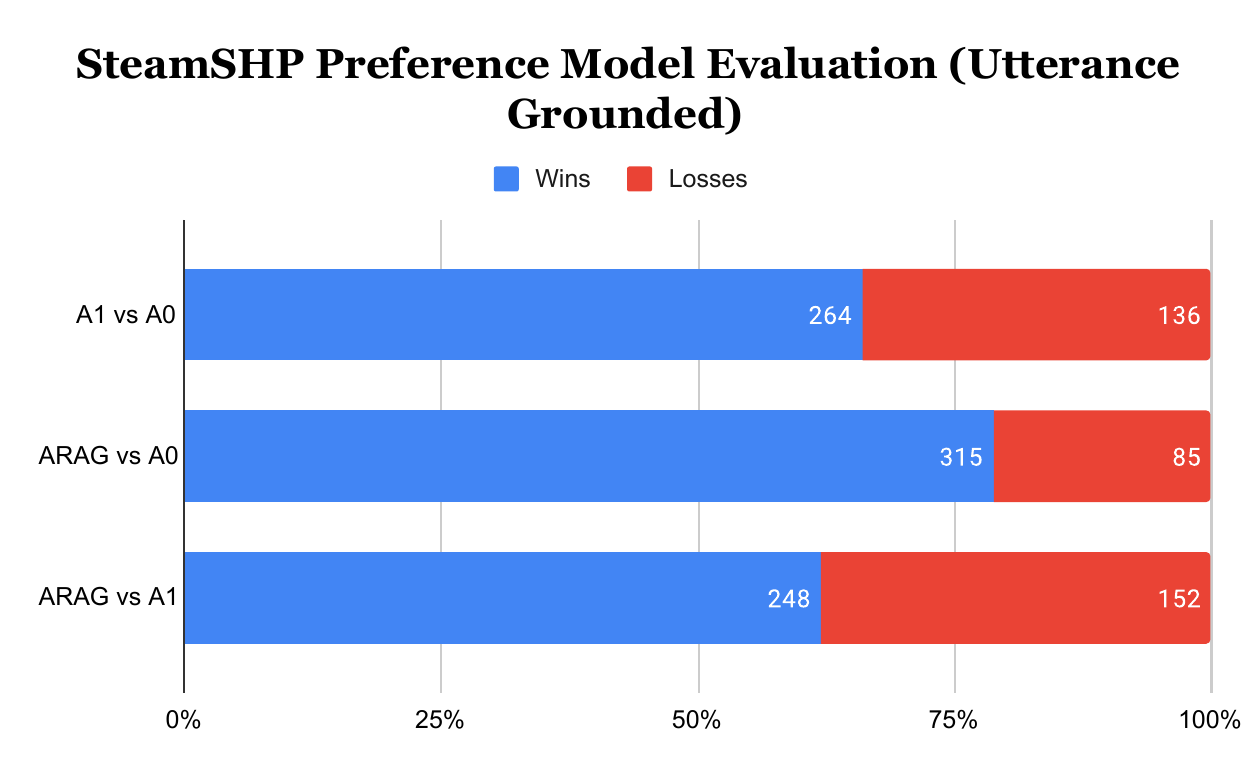}}
  \caption{SteamSHP Preference model Evaluation of Assistants created from Utterance Grounded Simulators against baseline assistant (A0)}
   \label{ug_sp}
\end{figure}
\begin{figure}[ht]
\centering
\fbox{
  \includegraphics[width=\columnwidth]{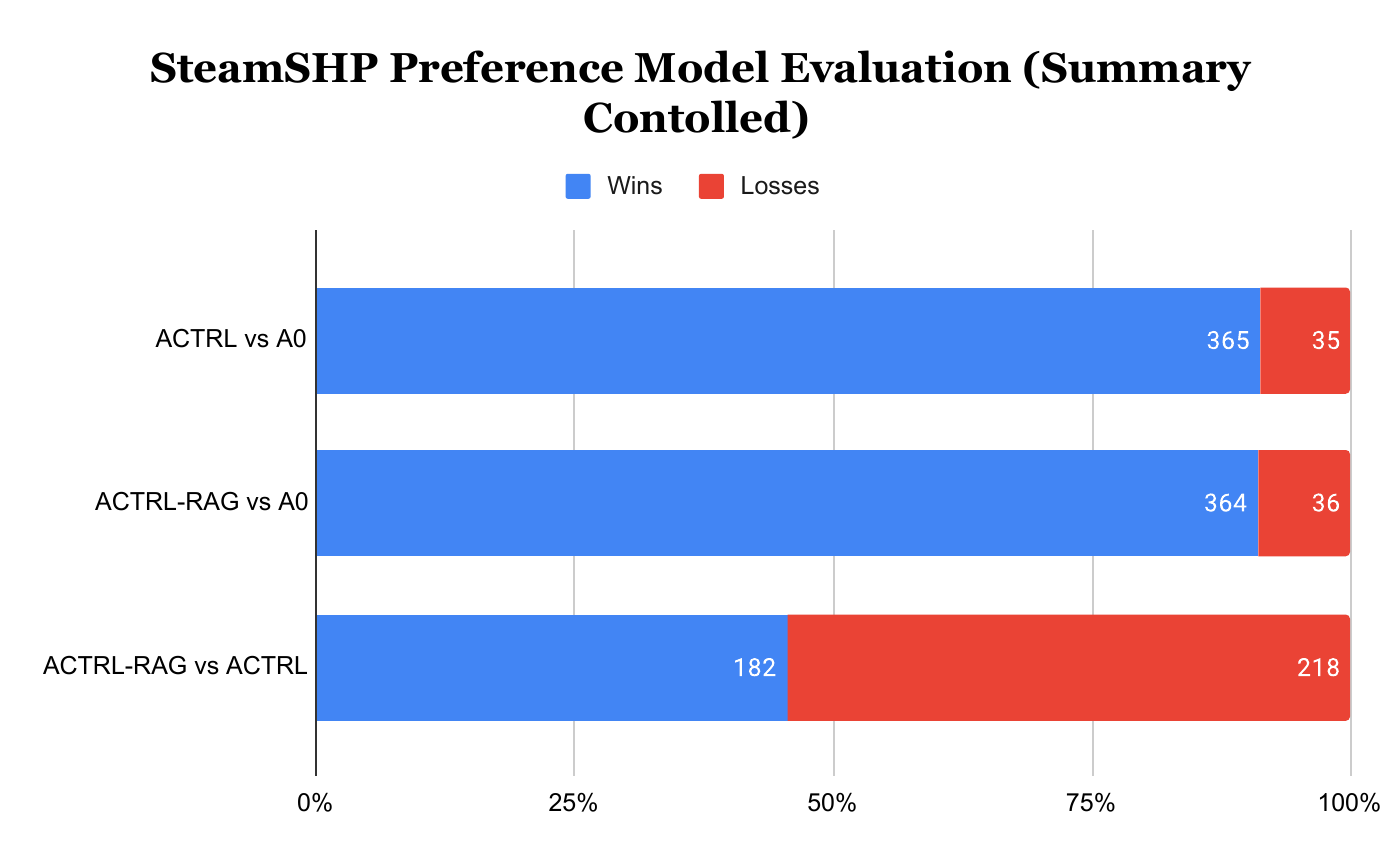}}
  \caption{SteamSHP Preference model Evaluation of Assistants created from Summary Controlled (-CTRL) against baseline assistant (A0)}
  \label{sg_sp}
\end{figure}
\textbf{Effect of Simulator:} We first compare A1 (i.e. the assistant trained on 8k interactions generated from S1) against A0 (i.e. the one trained without the help of the simulator). A1 outperforms A0 in all three evaluations as seen on the first rows of Figures~\ref{ug_fc},~\ref{ug_sr} and ~\ref{ug_sp}. The results are more prominent in SteamSHP’s evaluations. This shows that with the help of a simulator, we can generate more data and improve downstream assistant performance.

\textbf{Effect of Retrieval Augmentation:} We then compare whether an assistant model ARAG, trained from retrieval augmented data benefits training. With the retrieval augmented simulator, downstream performance across all metrics is improved. ARAG’s interactions are preferred more often as compared to A0 as well as A1 as seen in the 2nd and 3rd rows of Figures~\ref{ug_fc},~\ref{ug_sr} and ~\ref{ug_sp}.

\textbf{Effect of Summary Control:} The assistants ACTRL and ACTRL-RAG trained from the summary-controlled simulators are more often preferred across all the evaluations as shown in the first two rows of Figures~\ref{sg_fc},~\ref{sg_sr} and~\ref{sg_sp}. However, the non-retrieval counterpart ACTRL is more often preferred as compared to the retrieval counterpart.

\section{Conclusion}
Simulators provide a way to generate data to create downstream assistant models saving human time and effort. Through our framework~\Kaucus{}, we further showed that augmenting simulators by exploiting external knowledge helps generate diverse interactions and as well as creates more helpful assistants than vanilla simulators. We describe two types of knowledge-augmented simulators, a Retrieval Augmented Simulator, SRAG, and a summary-controlled simulator, SCTRL both of which consume external knowledge in unique ways.

Raw text is more prevalent than the conversational counterparts. Controlling simulators through conversational summaries or external documents can be a quick and powerful tool to convert public text to trainable interaction data and create more helpful assistants. It provisions the simulator to generate interactions for novel information outside the scope of an LLM's intrinsic parameters. We hope~\Kaucus{} will help encourage the development of automated techniques to be able to incorporate the vast amount of text produced rapidly over the internet and align assistant models better with newer data as well as be able to control the distribution of training data without the need for a rigid schema. 

\section*{Limitations}
Retrieval Augmentation helps incorporate diversity as well as benefit downstream models. We chose to use BM25 as our choice of retriever. However, there are dense retrievers~\cite{khattab2020colbert} and neural rerankers~\cite{pradeep2023rankzephyr} that perform better than BM25 across a range of information retrieval benchmarks. Our focus was to show the benefit of incorporating external knowledge while performing a rigorous set of experiments with the same. Future studies could specifically study the impact of additional hyper-parameter tuning by using varied choices of the retriever, the retrieving query, choice of summarisers and also gauge the impact of different domains than those of the Anthropic and the MSMarco datasets. 

Besides, our study does not consider the impact of prolonged training on generated data which could cause potential problems of model forgetting over the long run~\cite{shumailov2023curse}. More experiments conducted to gauge long-term viability would shed better light on the efficacy of knowledge simulators.

All the evaluations conducted in this paper were automated -- through popular reward or preference models. Human evaluations can provide better additional insights. Besides, the current intrinsic metrics primarily focus on diversity, which, while important, is only one dimension of dialogue evaluation and future work would benefit from other measures depending on the application.

\section*{Ethics Statement}
Our study has focused on the benefits of employing simulators to improve downstream assistant models. We believe that these simulators can also act as effective testers of assistants to pre-encounter and regurgitate harmful or undesirable assistant content before assistant models are deployed in impacting end applications. We should maintain caution against their unethical usage or if such regurgitation is exploited to cause harm. Just like assistants or other applications of large language models~\cite{dhole-2023-large}, simulators should also be gauged from a socio-technical lens, and appropriate checks and fallback mechanisms should be employed before their actual usage. Besides, simulators themselves could inadvertently learn biases in the training data, leading to unfair or biased generations, and can be exploited for malicious purposes such as generating fake news and harmful content or asking triggering questions.

\section*{Acknowledgements}
The author would like to thank the three anonymous reviewers for their useful suggestions.

\bibliography{anthology,custom}
\bibliographystyle{acl_natbib}

\end{document}